\documentclass[conference]{IEEEtran}
\IEEEoverridecommandlockouts
\usepackage{cite}
\usepackage{amsmath,amssymb,amsfonts}
\usepackage{algorithmic}
\usepackage{graphicx}
\usepackage{textcomp}
\usepackage{xcolor}
\usepackage{bm}
\usepackage[numbers]{natbib}
\usepackage{comment}

\def\BibTeX{{\rm B\kern-.05em{\sc i\kern-.025em b}\kern-.08em
    T\kern-.1667em\lower.7ex\hbox{E}\kern-.125emX}}
\begin{document}

\title{Financial Trading with Feature Preprocessing and Recurrent Reinforcement Learning 
}
\author{\IEEEauthorblockN{ Lin Li}
\IEEEauthorblockA{\textit{Department of Mathematics} \\
\textit{Southern University of Science and Technology}\\
Shenzhen, China \\
lilin@mails.ccnu.edu.cn}
}

\maketitle

\begin{abstract}
Financial trading aims to build profitable strategies to make wise investment decisions in the financial market. It has attracted interests in the machine learning community for a long time. This paper proposes to trade financial assets automatically using feature preprocessing skills and Recurrent Reinforcement Learning (RRL) algorithm. The strategy starts from technical indicators extracted from assets' market information. Then these technical indicators are preprocessed by Principal Component Analysis (PCA) and Discrete Wavelet Transform (DWT) and eventually inputted to the RRL algorithm to do the trading. The extensive empirical evidence shows that the proposed strategy is not only effective and robust in its performance, but also can mitigate the drawbacks underlying the initial trading using RRL.    
\end{abstract}


\section{Introduction}
With the fast development of machine learning in the last three decades, various machine learning algorithms have been developed and applied to automated trading in the financial market. 
The goal of financial trading is primarily to make profits while minimizing the risk of loss as much as possible. The key is, therefore, to make wise decisions based on fast-changing market conditions, such as taking a long position before the asset's price goes up while adopting a short position before the asset's price falls. 
One of the notable financial trading algorithms, handling this case, is RRL which was firstly proposed by \cite{moody1998performance,moody1998reinforcement,moody1999reinforcement,moody2001learning} and has been shown to be able to gain profit on various markets when tested on corresponding market data sets. However, it has also been shown to suffer from its own drawbacks. For instance, \cite{gold2003fx} tested RRL and its variants, two layer network RRL, on currency trading and reported that they can hardly find the optimal set of hyperparameters, due to there are a large number of hyperparameters that can only be tuned by trial and error and the interdependence amongst them. Moverover, \cite{gold2003fx} also reported the market dependence of the RRL performance. Besides, \cite{maringer2010threshold, maringer2012regime} argued that the original RRL lacks flexibility to trade in a regime-switching market where dramatic economic behavior could happen and the financial time series exhibits high non-linearity. They hence put forward the threshold and Regime-switching versions of the original RRL and tested them on artificial and real financial data sets. They argued that the out-of-sample results were generally support the regime-switching RRL, but there were still some doubts regarding the instability of the performance of the RRL based methods in the presence of transaction cost. Furthermore, \cite{molina2016stock} found that the basic RRL algorithm could hardly avoid precipitous drops in the stock price, which led to significant loss of cumulative profit, especially when stock prices did not exhibit structures.

In this paper, a strategy, combining the PCA and DWT operations to reduce dimensionality and noise for the feature set with RRL for financial trading, is presented. we aim to show how this strategy, named PCA\&DWT RRL, can greatly alleviate the above listed problems. Specifically, instead of executing the trading algorithm on the lagged return series which is adopted by most previous methods, PCA\&DWT RRL first extracts various technical indicators from each asset's market information, which are then preprocessed by PCA and DWT and eventually inputted to the RRL trading algorithm. We notice that other work related to trading using RRL approach, such as adding risk management layer and parameter optimization layer on top of the RRL trading layer \cite{dempster2006automated}, optimizing different objective functions \cite{2007Making} and others \cite{gold2003fx,maringer2010threshold,maringer2012regime}, primarily focused on designing involved algorithms to improve the original RRL performance. To the best of our knowledge, we are among the few work which tries to mitigate the drawbacks of the original RRL from the feature preprocessing's point of view. Feature preprocessing has been applied to different machine learning algorithms for handling financial data. For example, \cite{zhong2017forecasting} presented three dimensionality reduction methods, principal component analysis, fuzzy robust principal component analysis, and kernel-based PCA to reduce the dimension of the features before artificial neural networks were used to classify the direction of the stock daily return based on the prepocesed features data. Also, \cite{nobre2019combining} proposed to use PCA and DWT to preprocess the input data first, then the cleaned data was fed into the XG Boost machine to generate trading signals. Their trading actions are based on predictions of the price movement at the next period. Additionally, \cite{fengqian2020adaptive} considered the importance of feature selection in the process of making trading decisions. They first processed the financial data by K-line theory to produce candlesticks as an operation of denoising, then the candlesticks are decomposed into different subparts which are further clustered before put into the deep reinforcement learning model. However, in these cases, feature preprocessing is mainly used for forecasting securities' prices. In our case, accurate forecasting is neither necessary nor sufficient to a good performance of the RRL algorithm \cite{moody1998performance}. 

The main contributions of this paper are: to begin with, we propose to combine feature preprocessing operations, PCA and DWT, with basic RRL trading algorithm to trade in the financial market. Moreover, we solely adopt technical indicators extracted from each asset's market information as the raw feature instead of lagged return series of assets as adopted by many previous methods \cite{gold2003fx,maringer2012regime,dempster2006automated}. Additionally, we optimize Sharpe ratio directly with traditional batch update instead of other objectives with online learning \cite{moody2001learning}. Plus the well-developed packages for implementation of data processing, the proposed strategy is technically simpler than most previous methods based on RRL. Yet, it can attain impressive results. More importantly, the proposed strategy can greatly mitigate the issues with the basic RRL trading mentioned before. Last but not least, we present an overall empirical study of the proposed strategy and show that it is consistent, robust and reliable in most cases.  

\section{Data Preprocessing Module}

\subsection{Data Configuration}
In the financial market, price of stocks is affected by multiple factors which include but not limited to unexpected events, government policies and company activities. This implies that there may be a few periods when trading activities are not available, which results in a couple of \textit{NA} entries in the price time series. These \textit{NA}s are useless for our trading strategy and can be removed directly or filled with other values such as the last value ahead 
subject to specific learning task. In our case, we discard these \textit{NA} entries directly for the sake of simplicity. 
Additionally, technical analysis in the financial market has a long history amongst investment practitioners. As more and more evidence shows that markets are not as efficient as once believed \cite{LEBARON1999125}, technical analysis was applied to algorithmic trading \cite{dempster2001computational,nobre2019combining}. It is generally believed that technical analysis indicators can summarize the general pattern of the time series and mitigate local noise somehow in the data stream, which can further be utilized by the trading system to make profitable decisions \cite{dempster2001computational}. There are  quite  a  few  technical indicators developed by financial professionals \cite{achelis2001technical}. One can choose various technical indicators to use depending on specific tasks. While without enough number of indicators, it may be tough to reveal the pattern of the data stream comprehensively, including too many technical indicators may also affect the trading decision negatively and increase the computation burden, since the calculated value of some technical indicators are not always consistent with one another. Given this, the indicators used in this paper can be categorized into four groups, momentum indicators: Momentum (MOM), Moving Average Convergence Divergence (MACD), Money Flow Index (MFI), Relative Strength Index (RSI), volatility indicators: Average True Range (ATR), Normalized Average True Range (NATR), cycle indicators: Hilbert Transform Dominant Cycle Phase (HTDCP), Hilbert Transform Sinewave (HTS), Hilbert Transform Trend Market Mode (HTTMM) and volume indicators: Chaikin Oscillator (CO), On Balance Volume (OBV). Let $TI$ = \{MOM, MACD, MFI, RSI, ATR, NATR, HTDCP, HTS, HTTMM, CO, OBV\} denote the set of 11 technical indicators used for the following exposition.
The calculation of technical indicators was done via the python library TA-Lib \cite{talib}.
Additionally, to make the value of each feature on the same scale, we normalize each technical indicator stream with the standard normalization i.e. z-score which is represented by the following equation.
\begin{equation}
    X = \frac{X-\mu(X)}{\sigma(X)}
\end{equation}
where $X$ is the time series of each extracted feature, $\mu(X)$ and $\sigma(X)$ is the mean and standard deviation of $X$, respectively.

\subsection{Principal Component Analysis } 
The PCA operation of data preprocessing module receives the normalized feature set, including 11 technical indicators data streams, as input. The development of PCA originates from the curse of dimensionality which claims that data points in high dimension lie far away from each other, statistically speaking \cite{geron2019hands}. The curse of dimensionality not only makes the training of machine learning algorithm expensive, but also casts a shadow over the predictions of the trained algorithm, since data points in-sample and out-of-sample are so far away. In order to alleviate the effect of curse of dimensionality, a natural way is to reduce the dimension of data sets. That is where PCA comes into effect. To be more specific, PCA firstly fits the input, identifying the main components that represent the directions of maximum variance of the input. Then the main components are ordered according to the variance they explain and one can choose how many components are preserved. Afterwards, the original input is projected onto the retained components, resulting in a data set of lower dimension. In our case, the normalized technical indicators is a 11 dimensional data set at first. After being decomposed by PCA, there will be less than 11 indicators, reducing the probability of correlation and inconsistency amongst them. In this way, the processed set results in a new $TI^\prime \subset TI$. 
We use the well developed Scikit-learn \cite{pedregosa2011scikit} library to implement the PCA operation. Additionally, we set the hyperparameter, explained variance ratio, to 95\% which means that the sum of the variance explained by all retained principal components takes up at least 95\% of the total variance of the original data set.

\subsection{Discrete Wavelet Transform }
Although the data set returned by the PCA is simplified by removing less relevant features in the feature domain and possibly also reducing noise in each feature stream, some outliers or irrelevant data points, representing local noise, may still exist in each feature series. They may affect the training and trading of the RRL algorithm. To remove these local noise in the time domain of each technical indicator, we apply the discrete wavelet transform to the feature after being processed by PCA.

Reference \cite{mallat1989theory} proposed to calculate the DWT coefficients, including the approximation and the detail coefficients, using a pair of high pass and low-pass filter. In the algorithm, the father $\Phi(t)$ and mother $\Psi(t)$ basis functions are introduced to generate their corresponding son $\Phi_{j,k}(t)$ and daughter $\Psi_{j,k}(t)$ wavelets which are further utilized to approximate the original signal. 
The corresponding coefficients of son and daughter wavelets as a result of decomposing a function $f(x)$ is defined in the following inner product form \cite{mallat1989theory}. 
\begin{equation}
a_{j,k} = \langle f(t), \Phi_{j,k}(t) \rangle, \quad
d_{j,k} = \langle f(t), \Psi_{j,k}(t) \rangle
\end{equation}
where $a_{j,k}$ and $d_{j,k}$ are approximation and detail coefficients, respectively, and $k=0,1,2,\ldots$ and $j = 0,1,2,\ldots$. Though there are various types of wavelets, in this work, we use Haar wavelets with periodization padding mode, since Haar wavelets are useful to capture fluctuations between adjacent observations, recorded by \cite{Lahmiri2014Wavelet}, which would be heuristically useful to spot evident drawdowns in the financial market. Additionally, the decomposition can be iterated for multiple times, subject to the inherent decomposition level of the wavelets and the length of the signal or data series. 

Eventually, the DWT leaves us one set of the approximation coefficients and a couple of sets of the detail coefficients depending on the given decomposition level. In this paper, we set the DWT decomposition level equal to 4, since too high decomposition level would destroy the general pattern, while too low would still leave too much noise in the data \cite{lee2019pywavelets}. After the decomposition finishes, the general trend of the original signal is preserved in the approximation set, while the detail coefficients sets contain the local noise of the signal \cite{mallat1989theory}, which we aim to clean. 

At this point, we apply soft thresholding technique with an empirical threshold value equal to two times standard deviations of coefficients to each detail coefficients set.
Eventually, the inverse DWT method is used to reconstruct the signal which is the final denoised version of the original signal. By discarding the irrelevant coefficients, 
the reconstructed signal represents the essential characteristics of the original signal, which is further adopted by the RRL trader in the following trading module. 
The DWT process for each technical indicator series of each asset is implemented with the open source python package PyWavelets \cite{lee2019pywavelets}.

\section{Recurrent Reinforcement Learning for trading using technical indicators}
\subsection{Hypothesises}
In this paper, we make following assumptions which are also implied by RRL trading in most cases: the trader always takes either long or short position in the underlying asset, meaning that the trader is always in the market regardless of the market situation, although this may expose us to severe risk when the market volatility is high; the market is liquid enough such that trading orders can be immediately executed at the each day's close price of the underlying asset; the trading is put in a backtest environment. 
\subsection{Structure of RRL}
In this work, we make some modifications of the original RRL structure to make it more fit to our setting. Instead of optimizing differential Sharpe ratio \cite{moody1998performance} with online weight update, we directly optimize Sharpe ratio with the batch learning to update weights \cite{molina2016stock}. Moreover, instead of the one time training and test process adopted by \cite{moody1998performance,moody1999reinforcement,moody2001learning}, we adopt a rolling training and test process which will be illustrated further later. 
Given a sequence of the value of technical indicators in $TI^\prime$, the remained technical indicators after PCA and DWT operations, the trading position of RRL at each period is:
\begin{equation}\label{e1}
    F_t = F(\theta_t;{TI}^\prime,F_{t-1}) \in \{-1,1\}
\end{equation} 
where $F$ is the activation function, such as \text{sign} or \text{tanh}, $\theta_t$ is the set of internal system parameters, $F_{t-1}$ is the state of the position at the previous period. Here $F_{t-1}$ is added to the preprocessed feature set given the transaction cost incurred in the real trading process. The goal of the agent is to constantly optimize Sharpe ratio which for a time window $T$ is defined as:
\begin{equation}
    S_T = \frac{E[R_{1,\ldots,T}]}{\sqrt{E[R_{1,\ldots,T}^2] - (E[R_{1,\ldots,T}])^2}} = \frac{A}{\sqrt{B - A^2}}
\end{equation}
where $A = \frac{1}{T}\sum\limits_{t=1}^TR_t$ and $B = \frac{1}{T}\sum\limits_{t=1}^TR_t^2$. Note here we assume $R_t$ is the excess daily return over the risk free return of the investment for generality of discussion \cite{moody2001learning}. The daily profit of the trading strategy at period $t$, which is defined as:
\begin{equation} \label{e5}
    R_t = F_{t-1}\cdot r_t - \delta\cdot|F_t-F_{t-1}|
\end{equation}
This is the daily profit for $1$ share of the asset and $\delta$ is the transaction cost rate per share traded. The wisdom underlying (\ref{e5}) is that the trading algorithm rewards the action which is consistent with the following asset return, i.e. if one takes the long position ($F_{t-1}=1$) in advance and then the immediate asset return due to price movement is positive ($r_t>0$), then she gains profits, whereas the trading penalizes the action which is opposite to the price movement direction. The same mechanism holds for short positions. Given the nontrivial role of transaction cost in the process of active trading, especially high frequency trading, the daily profit of trading $R_t$ is dependent on the last time's trading position. If the trading positions are the same in two consecutive periods, there will be no transaction cost, while the transaction cost is doubled if the inverse is true. Moreover, the dependence of $R_t$ on the trading position at the last period makes the trading algorithm recurrent, because the last period's trading position $F_t$ is recursively dependent on $F_{t-1}$ according to (\ref{e1}). This is also why the algorithm is named RRL. Additionally, the cumulative profit of the trading system over a time window $T$ is:
\begin{equation}
    P_t = \sum\limits_{t=1}^TR_t = \sum\limits_{t=1}^T(F_{t-1}\cdot r_t-\delta \cdot |F_t-F_{t-1}|)
\end{equation}
Since we aim to maximize Sharpe ratio over time, gradient \textit{ascent} weight update method is employed, where the gradient of Sharpe ratio is calculated using the chain rule of derivatives as \cite{molina2016stock}:
\begin{equation}
    \begin{split}
    &\frac{dS_T}{d\theta_t}=\frac{dS_T}{dA}\frac{dA}{d\theta_t}+\frac{dS_T}{dB}\frac{dB}{d\theta_t}\\
    &=\sum\limits_{t=1}^T\Big\{\frac{dS_T}{dA}\frac{dA}{dR_t}+\frac{dS_T}{dB}\frac{dB}{dR_t}\Big\}\frac{dR_t}{d\theta_t}\\
    &=\sum\limits_{t=1}^T\Big\{\frac{dS_T}{dA}\frac{dA}{dR_t}+\frac{dS_T}{dB}\frac{dB}{dR_t}\Big\}\Big\{\frac{dR_t}{dF_t}\frac{dF_t}{d\theta_t}+\frac{dR_t}{dF_{t-1}}\frac{dF_{t-1}}{d\theta_t}\Big\}
    \end{split}
    \end{equation}
where terms $\frac{dS_T}{dA}$, $\frac{dS_T}{dB}$, $\frac{dA}{dR_t}$ and $\frac{dB}{dR_t}$ are calculated as:
\begin{align}
        \frac{dS_T}{dA} &= \frac{B}{(B-A^2)^{3/2}} &
        \frac{dS_T}{dB} &= -\frac{A}{2(B-A^2)^{3/2}}\\
        \frac{dA}{dR_t} &= \frac{1}{T}   &
        \frac{dB}{dR_t} &= \frac{2R_t}{T}
\end{align}
Other terms are calculated as follows \cite{molina2016stock}:
\begin{equation}
\begin{split}
    \frac{dR_t}{dF_t} &= \frac{d}{dF_t}(F_{t-1}\cdot r_t - \delta\cdot|F_t-F_{t-1}|) \\ &= -\delta \cdot \text{sign}(F_t-F_{t-1}) \\
        \frac{dR_t}{dF_{t-1}} &= \frac{d}{dF_{t-1}}(F_{t-1}\cdot r_t - \delta\cdot|F_t-F_{t-1}|) \\ &= r_t+\delta \cdot \text{sign}(F_t-F_{t-1})
    \end{split}
\end{equation}
Note that here we first discuss the sign of term $|F_t-F_{t-1}|$ to get rid of $|\cdot|$ which is not differentiable and then calculate the derivative of $R_t$ with respect to (w.r.t.) $F_t$. For the term $\frac{dF_t}{d\theta_t}$, we notice that if the decision function $F_t$ is in the form of \text{sign} function, it is not differentiable. One option would be to consider differentiable prethresholded output during training and discretize the output when trading \cite{moody2001learning}. For example, assuming $F_t =\text{tanh}(\theta_t^Tx_t)$, where $x_t = [1,{TI}_1^\prime,\ldots,{TI}_m^\prime,F_{t-1}]$ is the sequence of final features input to the trading system, and $m$ is the cardinality of ${TI}^\prime$. Then 
\begin{equation}
    \begin{split}
        \frac{dF_t}{d\theta_t} &= \frac{d}{\theta_t}\Big\{\text{tanh}(\theta_t^Tx_t)\Big\}
        = (1-\text{tanh}(\theta_t^Tx_t)^2)\cdot\frac{d}{d\theta_t}\Big\{\theta_t^Tx_t\Big\} \\
        &=(1-\text{tanh}(\theta_t^Tx_t)^2)\cdot\Big\{x_t+\theta_t^{M+2}\frac{dF_{t-1}}{d\theta_{t-1}}\Big\}
    \end{split}
\end{equation}
where $\theta_t^{M+2}$ is the is the $(M+2)\textsuperscript{th}$ parameter in the parameter vector $\theta_t$, corresponding to the entry $F_{t-1}$ in the feature set $x_t$, since the decision function $F_t$ is recurrently dependent on all its previous value. 

Once we obtain the gradient of Sharpe ratio $S_T$ w.r.t. algorithm parameters $\theta_t$, a batch gradient \textit{ascent} technique can be applied to update the algorithm parameters, i.e.
\begin{equation}
    \begin{split}
    \theta_{t+1} &= \theta_t + \rho \cdot (\frac{dS_T}{d\theta_t} - \lambda \cdot \theta_t) \\
   &= (1-\rho \cdot \lambda)\cdot \theta_t + \rho \cdot \frac{dS_T}{d\theta_t}
    \end{split}
\end{equation}
where $\rho$ and $\lambda$ are the learning rate and $l_2$ regularization parameter to avoid overfitting during training, respectively. Besides, we adopt a rolling training and trading techniques to implement the trading process of the proposed strategy. Specifically, the RRL trading agent starts from a set of initial parameters which are generated by a random seed, $\kappa$, then is trained on a training set of length $N_\text{train}$, then the system parameters are fixed to trade in the next $N_\text{trade}$ periods. Afterwards, the training and trading processes are repeated forward. The new training set would cover the previous test set and the new test set is also advanced forward, following the new training set. This process is repeated until the last batch of trading periods. During training, the agent is trained for $n$ epochs and the process is early stopped if the objective value in two consecutive epochs is less than a threshold, $\epsilon$. This is also to avoid overfitting of the RRL algorithm and ensure a better generalization ability of the algorithm. 

\section{Experiments}
\subsection{Data sets} 
We test the proposed trading strategy on real data sets from different financial markets. To be more specific, 3 data sets are collected and each consists of the daily prices (Open, High, Low, Close) and Volume over the period of 01/01/2001 to 12/03/2021. Moreover, in order to test the robustness of the proposed trading strategy in different financial markets, we allow each data series to exhibit a unique pattern to ensure the diversity of the data sets. The experiments are carried out in the following financial markets: NYSE Composite index (NYSE), Exxon Mobile Corporation stocks (XOM) and Corn futures contract (ZCF). XOM and ZCF are also selected as data sets in \cite{nobre2019combining}, although in different time frame. Data points of NYSE, XOM and ZCF are downloaded from Yahoo Finance \footnote{accessible from https://finance.yahoo.com/.}. The close price series of each data set is shown in the first row in Fig. \ref{f1}. Clearly, each price path is rough and full of ups and downs, including precipitous drops.  
\subsection{Performance Metrics}
The metrics used to measure the performance of different trading strategies, including PCA\&DWT RRL, in the real financial market are: Net Profit (NP) which is final wealth $P_T$ accumulated by the RRL trading over horizon $T$ in the RRL setting; Annualized Percentage Yield (APY), representing the annualized percentage gain; Annualized Sharpe Ratio (ASR), representing the annualized risk adjusted return and we assume the risk-free return is 1\% per annum which is approximate to the US 10 year treasury bond yield as of year 2020 and there are 252 trading days each year; Maximum Drawdown (MDD) which measures the profit decline percentage from peak value (PV) before largest drop and lowest value (LV) before new high established of an investment during a specific period; Calmar Ratio (CR), indicating the level of risk taken to achieve a return and a higher CR suggests that the strategy's return is not at the risk of large drawdowns and vice verse. The higher the numerical value of these metrics are, the better the performance of the trading strategy is except for MDD for which a lower value is preferred, since most investors are risk-averse. The definitions for these metrics are summarized in Table \ref{t2}.
 \begin{table}[t]        
    \caption{Metrics used to measure the performance of different methods}
     \label{t2}
     \resizebox{\columnwidth}{!}{
        \begin{tabular}{ccccc}
            \hline
             NP &  APY & ASR &  MDD & CR  \\  
             \hline
             $P_T$ & $(\frac{P_T+{MP}_0^{\mathrm{a}}}{{MP}_0})^{\frac{252}{T}} - 1$  & $\frac{\text{Mean}(R_{1,\ldots,T}) - 0.01}{\text{Std}(R_{1,\ldots,T})}\cdot \sqrt{252}$  & $\frac{\text{PV}-\text{LV}}{\text{PV}}$ & $\frac{\text{APY}}{\text{MDD}}$ \\
            \hline
            \multicolumn{4}{l}{$^{\mathrm{a}}$Market close price of the asset at the beginning of trading.}
        \end{tabular}
         }
    \end{table}

\subsection{Benchmark Strategies}
A way to show the effectiveness of the proposed trading strategy is to compare the performance of the strategy with other benchmark strategies. In this paper, we compare the proposed trading method with the baselines: Buy and Hold (B\&H), which means to maintain a long only position for the whole trading periods; an active trading strategies based on autoregressive integrated moving average model (ARIMA), which firstly predicts the price of the asset at the next period and then takes a long position if the return calculated from prediction is positive or takes a short position if the inverse is true \cite{Kumar}. In order to evaluate the effect of feature preprocessing using PCA and DWT techniques, we also present the results of the proposed method without PCA and DWT operations (TI RRL). Moreover, to delve deep into the importance of each feature preprocessing part, performance of method with feature only processed by PCA (PCA RRL) and method with feature only processed by DWT (DWT RRL) is tested. Additionally, since PCA can also remove noise in the data sets, which is somehow overlapped with the use of DWT, we will show the performance of the method with feature firstly processed by PCA and then DWT (the proposed PCA\&DWT RRL) and the method whose feature firstly processed by DWT and then PCA (DWT\&PCA RRL). We want to have an empirical impression on the effect of feature processing order on trading performance. 

\subsection{Hyperparameters}
A common feature of most machine learning models is that their performance depends on the setting of hyperparameters which are parameters set before the training process begins. In our case, besides the hyparameters described before in the data preprocessing module, we empirically fix $N_\text{train}=500, \rho=0.1, \lambda=0.01, n=100, \kappa=42, \epsilon=0, N_\text{trade}=500$ the same for all 3 data sets except for transaction cost rate, $\delta$, which has a practical meaning. Refering to \cite{nobre2019combining}, we decide to set $\delta$ equal to $1$\$ per share traded, $0.01$\$ per share traded and $0.001$\$ of the future contract spot price per contract traded for data sets, NYSE, XOM and ZCF, respectively. 

\subsection{Numerical Results}
\subsubsection{General performance of PCA\&DWT RRL}
\begin{figure*}[htbp]
\includegraphics[width=\textwidth]{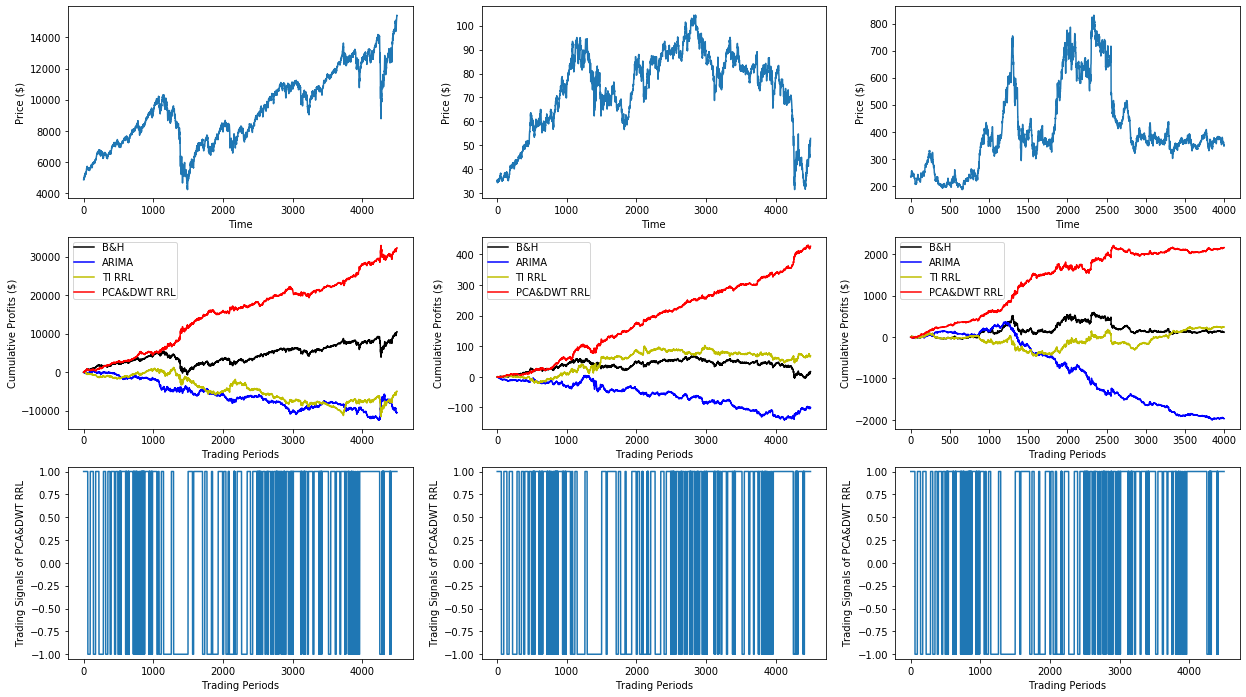}
\caption{The top row shows the close price trends of data sets NYSE, XOM and ZCF. The middle row shows the cumulative sum of the Net Profit (NP) achieved by strategies B\&H, ARIMA, TI RRL, PCA\&DWT RRL over the horizon. The bottom row shows the trading signals with frequency 518, 729, 488 times in total from left to right given by the PCA\&DWT RRL during trading process. }
\label{f1}
\end{figure*}
\begin{table}
  \caption{Numerical value of the performance of strategies B\&H, ARIMA, TI RRL, PCA\&DWT RRL on different data sets}
  \label{t3}
  \resizebox{\columnwidth}{!}{
  \begin{tabular}{cccccl}
    \hline
    Data sets & Metrics & B\&H & ARIMA & TI RRL & PCA\&DWT RRL\\
    \hline
    NYSE & NP(\$) & 10504.62 & -10517.95 & -4918.31 & \textbf{32286.96} \\
         & APY & 0.07 & -2.07 & -2.04 & \textbf{0.12} \\
         & ASR & 0.35 & -0.35 & -0.17 & \textbf{1.07} \\
         & MDD & 1.12 & $40.41^{\mathrm{a}}$ & 10.35 & \textbf{0.12} \\
         & CR & 0.06 & -0.05 & -0.2 & \textbf{1.05} \\
         \hline
    XOM & NP(\$) & 17.15 & -100.35 & 68.78 & \textbf{426.06} \\
        & APY  &  0.02  &  -2.08  & 0.06  & \textbf{0.16}  \\
        & ASR  &  -0.09 & -0.47 & 0.08 & \textbf{1.23} \\
        & MDD & 1.06 & 31.30 & 0.56 & \textbf{0.28}  \\
        & CR & 0.02 & -0.07 & 0.11 & \textbf{0.55} \\
        \hline
    ZOF &NP(\$)  & 114.00 & -1971.00 & 237.25 & \textbf{2158.61} \\
        & APY  &  0.03 & -2.15 & 0.04 & \textbf{0.16} \\
        & ASR & 0.03 & -0.91 & 0.09 & \textbf{0.96} \\
        & MDD & 0.89 & 6.41 & 5.83 & \textbf{0.15} \\
        & CR & 0.03 & -0.34 & 0.01  & \textbf{1.04} \\
  \hline
  \multicolumn{4}{l}{$^{\mathrm{a}}$The MDD could be greater than 1 or 100\%, if the LV is near 0.}
\end{tabular}
}
\end{table}
\begin{figure}
    \centering
    \includegraphics[width=\columnwidth]{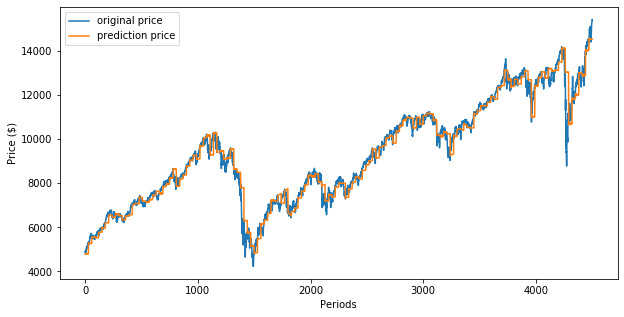}
    \caption{Prediction of NYSE price given by ARIMA(3,1,2) model. }
    \label{f2}
\end{figure}
In this section, we evaluate the performance of the proposed trading strategy on selected data sets. Specifically, the numerical value of different metrics achieved by execution of strategies, B\&H, ARIMA, TI RRL and PCA\&DWT RRL on data sets, NYSE, XOM and ZOF are summarized in Table \ref{t3} with the best result regarding each metric stressed in bold face. It takes us around 26.64 seconds, 26.86 seconds and 23.64 seconds on a 4 core Intel i5 PC for the propsed strategy to complete running on NYSE, XOM and ZCF data sets, respectively. Note that TI RRL is just a simplified version of PCA\&DWT RRL without feature preprocessing operations, PCA and DWT. Especially, the common hyperparameters of them are the same. From Table \ref{t3}, it is clear that PCA\&DWT RRL significatly outperforms all baseline strategies, including TI RRL on all three data sets, i.e. three different markets, in terms of all listed metrics, meaning that feature preprocessing steps are necessary in boosting the performance of TI RRL. Moreover, from Fig. \ref{f1}, it is obvious that the proposed strategy can effectively avoid large drawdowns and eventually achieve impressive cumulative sum of net profit on all data sets via making smart decisions on investment positions, i.e. long or short, during different trading periods. For instance, on the NYSE data set, the PCA\&DWT RRL strategy successfully avoids large drawdown of profits at periods around 1500 and 4300 by taking short positions. While the proposed strategy may not able to make correct decisions at every period for each data set, we argue that this strategy makes correct decisions in most cases. Besides, we attain these results with the same set of hyperparameters, meaning that our strategy does not need to frequently tune hyperparameters to obtain a good result in different markets, even under transaction costs. This clearly mitigates the issues, market-dependent performance, instable performance of RRL based trading and loss of profits due to price precipitous drops, raised in \cite{gold2003fx,maringer2010threshold,maringer2012regime,molina2016stock}. 

As to the ARIMA strategy, in this work, we first find a suitable ARIMA model on each data set separately \cite{Kumar}, which leads us to ARIMA(3,1,2) on NYSE, ARIMA(2,1,0) on XOM and ARIMA(1,1,1) on ZCF, respectively, then we find the optimal model parameters on a training window of length 500 periods and predict the future price for another window of length 30 periods. This fitting and prediction process are repeated forward until the last batch of prediction. The prediction result for NYSE data set, for example, is presented in Fig. \ref{f2}, revealing that although ARIMA(3,1,2) can predict the general price trend, there are inevitable time lags between the true price value and the prediction one, due to the model characteristics. We argue that these lagged prediction prices result in the worst performance of the ARIMA based trading strategy given in Table \ref{t3}, given its relatively simple trading logic. On the other hand, although the B\&H strategy is a passive investment strategy which is completely determined by the price movement of each asset, Table \ref{t3} shows that it can substantially outperform ARIMA strategy and sometimes even beat TI RRL on the NYSE data set, meaning following the market is normally a viable strategy.

\subsubsection{Effect of PCA and DWT}
\begin{table}
  \caption{Numerical value of the performance of strategies TI RRL, PCA RRL, DWT RRL and PCA\&DWT RRL on different data sets}
  \label{t4}
  \resizebox{\columnwidth}{!}{
  \begin{tabular}{cccccl}
    \hline
    Data sets & Metrics &  TI RRL & PCA RRL & DWT RRL & PCA\&DWT RRL\\
    \hline
    NYSE & NP(\$) & -4918.31& -2748.91 & 24619.60 & \textbf{32286.96} \\
         & APY & -2.04 & -2.03 & 0.11 & \textbf{0.12} \\
         & ASR & -0.17& -0.09 & 0.82 & \textbf{1.07} \\
         & MDD & 10.35& 12.05& 0.20 & \textbf{0.12} \\
         & CR & -0.20 & -0.17 & 0.53 & \textbf{1.05} \\
         \hline
    XOM & NP(\$) & 68.78 &3.38&413.02& \textbf{426.06} \\
        & APY  & 0.06& 0.01 & 0.15  & \textbf{0.16}  \\
        & ASR  & 0.08 &-0.13 & 1.19 & \textbf{1.23} \\
        & MDD & 0.56 &0.96 & 0.34& \textbf{0.28}  \\
        & CR & 0.11 & 0.01 & 0.45& \textbf{0.55} \\
        \hline
    ZCF &NP(\$)  & 237.25 & -158.77 & 2084.75 & \textbf{2158.61} \\
        & APY  & 0.04& -2.03 & \textbf{0.16} & \textbf{0.16} \\
        & ASR &0.09 & -0.09 & 0.93 & \textbf{0.96} \\
        & MDD & 5.83 & 200.19 & 0.16 & \textbf{0.15} \\
        & CR & 0.01 &-0.01 & 0.99  & \textbf{1.04} \\
  \hline
\end{tabular}
}
\end{table}
The performance of PCA RRL and DWT RRL together with the basic TI RRL and the proposed strategy is presented in Table \ref{t4}. By examining the table, it is easy to find that TI RRL and PCA RRL behave worse than the other two strategies. By comparing PCA RRL with PCA\&DWT RRL, one could find that DWT technique can substantially improve the performance of PCA RRL, meaning that removing noise in the data plays an important role in the proposed strategy. Comparing PCA RRL with DWT RRL shows that DWT technique is more important to boost the basic TI RRL's performance, whereas only execution PCA technique is not able to improve the basic TI RRL's performance effectively. However, PCA technique does play a positive role in the proposed strategy since PCA\&DWT RRL is slightly outperforms the DWT RRL strategy in all data sets regarding most metrics.   
Furthermore, to evaluate the effect of the order of the feature preprocessing techniques on trading performance, we further test the performance of PCA\&DWT RRL and DWT\&PCA RRL on the three data sets. The numerical value of their performance is listed in Table \ref{t5}, which exhibits that the proposed strategy outperforms the DWT\&PCA RRL on NYSE and XOM data sets, while DWT\&PCA RRL strategy outperforms the proposed strategy on ZCF data set. Therefore, although there are some differences between the performance of these two strategies, it is tough to conclude which one is definitely better than the other. The effect of the order of feature preprocessing to the RRL based trading is market-dependent.
\begin{table}
  \caption{Numerical value of the performance of strategies DWT\&PCA RRL and PCA\&DWT RRL on different data sets}
  \label{t5}
  \begin{tabular}{cccccl}
    \hline
    Data sets & Metrics & DWT\&PCA RRL & PCA\&DWT RRL\\
    \hline
    NYSE & NP(\$) & 23848.53 & \textbf{32286.96} \\
         & APY & 0.10 & \textbf{0.12} \\
         & ASR & 0.79 & \textbf{1.07} \\
         & MDD & 0.20 & \textbf{0.12} \\
         & CR & 0.52 & \textbf{1.05} \\
         \hline
    XOM & NP(\$) & 276.24 & \textbf{426.06} \\
        & APY  & 0.13  & \textbf{0.16}  \\
        & ASR  & 0.75 & \textbf{1.23} \\
        & MDD & 12.68 & \textbf{0.28}  \\
        & CR & 0.01 & \textbf{0.55} \\
        \hline
    ZCF &NP(\$)  & \textbf{2303.39} & 2158.61 \\
        & APY  & \textbf{0.16} & \textbf{0.16} \\
        & ASR & \textbf{1.03} & 0.96 \\
        & MDD & \textbf{0.07} & 0.15 \\
        & CR & \textbf{2.32} & 1.04 \\
  \hline
\end{tabular}
\end{table}
\subsubsection{Hyperparameter sensitivity}
\begin{figure*}[htbp]
\includegraphics[width=\textwidth]{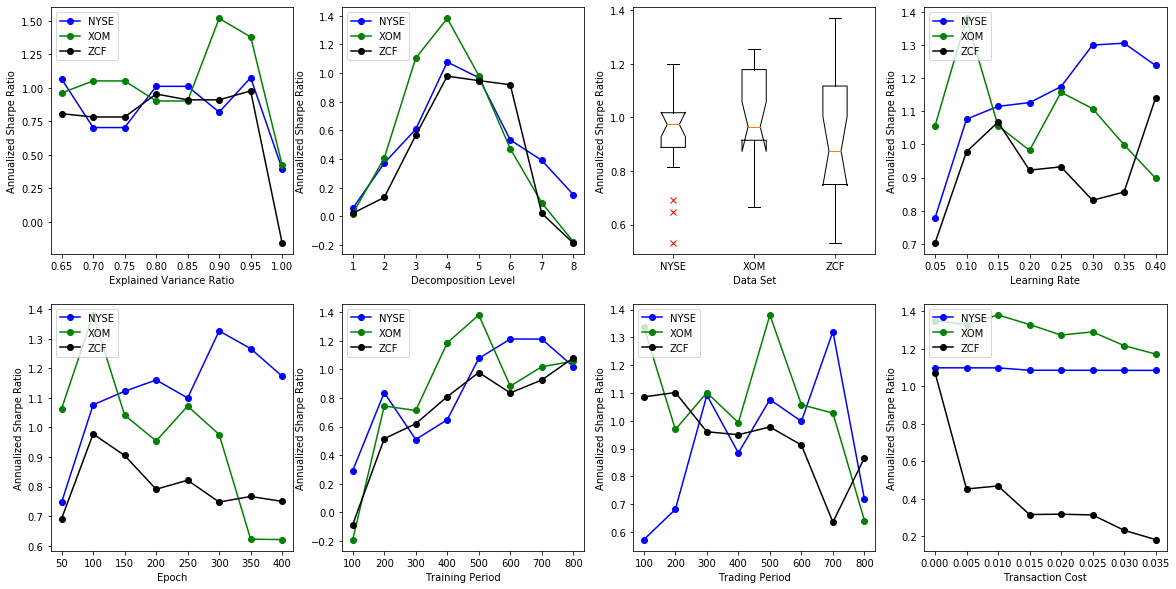}
\caption{From top to bottom and left to right, the Annualized Sharpe Ratio achieved by the PCA\&DWT RRL strategy w.r.t. various hyperparameters, i.e. explained variance ratio in the PCA part, decomposition level in the DWT part, the initial parameter, learning rate, training epoch, training period, trading period and transaction cost in the RRL part, respectively, on NYSE, XOM and ZCF data sets. The boxplot on the first row represents the result with twenty different initial parameters of the strategy on each data set. }
\label{f3}
\end{figure*}
As is well known, in many machine learning algorithms, hyperparameters directly determine the performance of the algorithm since they control the final parameters the algorithm finds, which decides the generalization ability and robustness of the algorithm. Although the proposed algorithm is able to gain superior results in different markets as shown in Table \ref{t3} with the same set of hyperparameters, which has arguably verified its out-of-sample robustness, we further test the sensitivity of this algorithm to its hyperparameters for more general settings. The result is shown in Fig. \ref{f3}, where in each graph, the respective hyperparameter is tuned with other hyperparameters fixed. By Fig. \ref{f3}, it is clear that the proposed strategy maintains a stable performance within a reasonable range when the respective hyperparameter value varies in the given range. For instance, the Annualized Sharpe Ratio obtained by the proposed strategy w.r.t. explained variance ratio fluctuates around 0.75 on different data sets except the 1.00 point at the x-label. The transaction costs listed at x-label of the last plot in Fig. \ref{f3} are particularly high for ZCF market in comparison to the normal transaction cost, 0.001 of per traded contract value. Even so, the proposed strategy achieves positive ASR value. In summary, the PCA\&DWT RRL strategy is able to maintain a robust performance w.r.t. its hyperparameters.    

\section{Conclusion}
This paper proposes to combine feature preprocessing and RRL to trade in the financial market. The feature preprocessing operations, represented by PCA and DWT, reduce the dimension and noise in the technical indicator feature sets. The processed feature is eventually inputted to the RRL algorithm to make trading decisions in the market, which leads to promising results. Moreover, the robustness and consistency in the performance of the proposed strategy clearly mitigate the drawbacks of the original RRL trading algorithm. Future work could be directed to analyse in detail which part i.e. optimized objective, training and trading method or feature preprocessing leads to the improvement of the performance of the RRL algorithm. 

\section*{Acknowledgment}
I would like to thank Jiawen Gu, Hao Ni and Jinsong Zheng for valuable support of this project.

\bibliographystyle{IEEEtranN}
\bibliography{mybibfile}
\end{document}